\documentclass{article}
\usepackage{spconf}
\usepackage{amsmath,graphicx}

\usepackage{booktabs}
\usepackage{graphicx}
\usepackage[table,xcdraw]{xcolor}
\usepackage{fontawesome}
\usepackage{multirow}
\usepackage{url}
\usepackage{dblfloatfix}
\usepackage{textcomp}

\newcommand{\unipa}{$^{\star}$}
\newcommand{\enna}{$^{\dagger}$}
\newcommand{\poli}{$^{\ddagger}$}

\title{Benchmarking Representations for Speech, Music, and Acoustic Events}
\name{
    Moreno La Quatra\enna, 
    Alkis Koudounas\poli, 
    Lorenzo Vaiani\poli \\
    \textit{Elena Baralis}\poli, 
    \textit{Luca Cagliero}\poli, 
    \textit{Paolo Garza}\poli, 
    \textit{Sabato Marco Siniscalchi}\unipa
}
\vspace{5mm}
\address{
    {\enna Kore University of Enna}, 
    {\poli Politecnico di Torino},
    {\unipa Università degli Studi di Palermo}
}
\begin{document}

\maketitle

\begin{abstract}
Limited diversity in standardized benchmarks for evaluating audio representation learning (ARL) methods may hinder systematic comparison of current methods' capabilities. We present ARCH, a comprehensive benchmark for evaluating ARL methods on diverse audio classification domains, 
covering acoustic events, music, and speech.
ARCH comprises 12 datasets, that allow us to thoroughly assess pre-trained SSL models of different sizes.
ARCH streamlines benchmarking of ARL techniques through its unified access to a wide range of domains and its ability to readily incorporate new datasets and models.
To address the current lack of open-source, pre-trained models for non-speech audio, we also release new pre-trained models that demonstrate strong performance on non-speech datasets.
We argue that the presented wide-ranging evaluation provides valuable insights into state-of-the-art ARL methods, and is useful to pinpoint promising research directions.

\end{abstract}
\section{Introduction}
\label{sec:intro}

Audio representation learning (ARL) has emerged as a promising research area for the design of general-purpose architectures for audio processing.
The goal of ARL is to encode audio signals into meaningful high-level feature representations that can be used for a wide range of downstream tasks, such as automatic speech recognition (ASR)~\cite{w2v2}, music information retrieval (MIR)~\cite{castellon2021calm}, and acoustic event detection (AED)~\cite{wu2022efficacy}.
Recently, self-supervised pre-training approaches, e.g.,  Wav2Vec 2.0~\cite{w2v2} and HuBERT~\cite{hubert}, have demonstrated strong performance on speech tasks by pre-training transformer-based architectures on large amounts of audio data.
However, the performance of these models on non-speech audio tasks has not been extensively evaluated, and the lack of open-source pre-trained models for comprehensive audio processing hinders the development of ARL methods in domains beyond speech.

The growing demand for effective, general-purpose audio processing techniques has led to the development of various ARL models~\cite{pascual19_interspeech, pretext_multitask, BYOL23}. 
Evaluation frameworks are crucial for measuring models' performance, and several benchmarks have been proposed, including HARES~\cite{HARES}, SUPERB~\cite{SUPERB}, LeBenchmark~\cite{LeBenchmark}, LAPE~\cite{LAPE}, and HEAR~\cite{HEAR}. 
However, existing benchmarks are limited in scope, accessibility, or flexibility.
HARES~\cite{HARES} mainly focuses on audio representation extracted from spectrograms and has limited open-source access for evaluation. 
LAPE~\cite{LAPE} provides valuable additional perspectives by focusing on pre-training and evaluating self-supervised (SSL) models, with an emphasis on low-resource settings. 
SUPERB~\cite{SUPERB} and LeBenchmark~\cite{LeBenchmark} are two complementary benchmarks that evaluate the performance of SSL models on speech-related tasks.
HEAR~\cite{HEAR} is a benchmark designed to evaluate general-purpose ARL models across different task domains. 

In this paper, we introduce ARCH (i.e., Audio Representation benCHmark), a comprehensive framework for evaluating ARL methods on diverse audio classification domains.
It allows standardized evaluation and comparison of ARL models on a wide range of domains, including acoustic events, music, and speech.
SUPERB and LeBenchmark differ in scope from ARCH, focusing on the speech domain rather than the diverse range of audio tasks.
HEAR, on the other hand, has goals similar to ARCH in benchmarking representation models across different audio domains.
However, the two benchmarks can be seen as complementary resources for the following reasons:
(1) They have minimal dataset overlap (i.e., only two datasets in common), providing complementary evaluation perspectives.
(2) HEAR is designed as a fixed benchmark,
whereas ARCH is readily extensible \textit{by design} to incorporate new datasets and models. 
Providing an additional evaluation resource, ARCH applies an alternate approach for fully benchmarking audio representation learning models.

This paper aims to advance audio representation learning through three main contributions.
First, we introduce a new extensible framework for the standardized evaluation of ARL models across diverse domains.
Second, addressing the lack of open-source pre-trained models for non-speech tasks, we also release general-purpose models demonstrating strong performance in acoustic events and music domains.
Third, we present an extensive comparative study of state-of-the-art methods evaluated on ARCH to provide valuable insights while highlighting promising research directions.

\section{The ARCH benchmark}

\begin{table}[]
\caption{
Datasets included in ARCH with their corresponding domains (acoustic events (\faVolumeDown), music (\faMusic), and speech (\faUser)), classification task types (single S or multi-label M), number of samples, average duration, and number of classes.
}
\vspace{0.5em}
\label{tab:data_arch}
\resizebox{\columnwidth}{!}{%
\begin{tabular}{@{}lccccc@{}}
\toprule
Dataset & Domain & Task & Samples & Avg duration & Classes \\ \midrule
ESC-50~\cite{ESC50} & \faVolumeDown & S & 2000 & 5.0s & 50 \\
US8K~\cite{US8K} & \faVolumeDown & S & 8732 & 3.61s & 10 \\
FSD50K~\cite{FSD50K} & \faVolumeDown & M & 51197 & 7.64s & 200 \\
VIVAE~\cite{VIVAE} & \faVolumeDown & S & 1085 & 0.90s & 6 \\
FMA~\cite{FMA} & \faMusic & S & 8000 & 29.98s & 8 \\
MTT~\cite{MTT} & \faMusic & M & 21108 & 29.12s & 50 \\
IRMAS~\cite{IRMAS} & \faMusic & M & 8278 & 5.73s & 11 \\
MS-DB~\cite{MSDB} & \faMusic & S & 21571 & 2.97s & 8 \\
RAVDESS~\cite{RAVDESS} & \faUser & S & 1440 & 3.70s & 8 \\
AM~\cite{AudioMNIST} & \faUser & S & 30000 & 0.64s & 10 \\
SLURP~\cite{SLURP} & \faUser & S & 72396 & 2.85s & 77 \\
EMOVO~\cite{EMOVO} & \faUser & S & 588 & 3.12s & 7 \\ \bottomrule
\end{tabular}%
}
\end{table}

This section outlines the modular framework design and standardized evaluation procedure implemented in ARCH\footnote{\url{https://github.com/MorenoLaQuatra/ARCH}}.

\subsection{Framework Architecture}
\label{ssec:arch_architecture}
ARCH employs a modular architecture to enable easy integration of new datasets and models in the benchmark.
The core framework is implemented in Python and provides a simple interface for benchmarking audio representation models. New datasets can be added by creating a distinct class that handles loading collection-specific information.
Any required data pre-processing and metadata are encapsulated in this class.
It also encapsulates the logic to iterate batches and evaluate embeddings on its target task. 
Training and evaluation loops are standardized for both single- and multi-label classification tasks. 
Only minimal changes are required to adapt the process to new datasets, such as modifying the data splitter. 
This enables customization for different data while keeping a common interface and protocol. 
Adding a new model instead entails creating a model wrapper.
It exposes a specific method that handles generating sample-level embeddings from the raw audio. 
The model-specific logic is abstracted from the benchmarking workflow.

\subsection{Evaluation procedure}
\label{sec:evaluation_procedure}

We follow a standardized evaluation process to assess the quality of learned audio representations.
Since the goal is to evaluate the representations in themselves, rather than optimizing classification performance, the process does not allow model fine-tuning.
The evaluation process implements a consistent protocol across all datasets for fair comparison.
All models evaluated in this paper generate frame-level vector representations.
To obtain a single embedding vector representing the entire audio sample, we perform average pooling on the sequence of frame-level representations.
This fixed-dimensional vector is then fed into a single linear layer,  which is trained for 200 epochs with the AdamW optimizer~\cite{adamw} to classify the embeddings.
The learning rate scheduler applies a linear warmup for the first 10\% of steps, increasing the learning rate from zero to reach a maximum value of 0.001.
This is followed by linear decay for the remainder of training.
Applying a  simple linear classifier we can assess the intrinsic quality of the representations, excluding the effects of additional nonlinear processing.

The modular design of ARCH allows easy integration of new models through specific wrappers that are used to extract sample representations.
However, we explicitly avoid methods leveraging additional trainable parameters to prevent discrepancies in the evaluation process (e.g., Attention Pooling).
This choice may limit the final accuracy of the models but ensures a fair comparison of the learned representations.
The key goal of the proposed evaluation procedure is to devise an objective evaluation that can guide the selection of optimal models for specific tasks or domains based only on their inherent capability to capture relevant information.

\section{Datasets and Models}
\label{sec:datasets}

ARCH includes 12 datasets spanning three domains: acoustic events, music, and speech, each of them covered by four publicly available data collections to ensure reproducibility.
Table~\ref{tab:data_arch} details information about the datasets.

The acoustic events data are collected from the following datasets: ESC-50~\cite{ESC50}, UrbanSound 8K (\texttt{US8K})~\cite{US8K}, FreeSound Dataset 50K (\texttt{FSD50K})~\cite{FSD50K}, and Variably Intense Vocalizations of Affect and Emotion (\texttt{VIVAE})~\cite{VIVAE}. These datasets provide a good coverage of environmental sounds, urban sounds, and human vocalizations. 
Three out of four acoustic events datasets involve single-label classification, while FSD50K employs multi-labeling (i.e., each audio sample can be tagged with multiple categories).
Sample lengths vary between roughly 1 and 7.5 seconds.

The datasets in the music domain include Free Music Archive (\texttt{FMA})~\cite{FMA}, MagnaTagATune (\texttt{MTT})~\cite{MTT}, Instrument Recognition in Musical Audio Signals (\texttt{IRMAS})~\cite{IRMAS}, and Medley-solos-DB (\texttt{MS-DB})~\cite{MSDB}.
Such a selection enables the evaluation of tasks like genre classification, tagging, and instrument recognition.
Furthermore, two of the datasets involve multi-label classification (i.e., \texttt{MTT} and \texttt{IRMAS}), whereas two are single-label tasks. 
The average sample duration varies considerably, ranging from 3 to 30 seconds.

The literature on the speech domain is quite rich and includes many publicly available datasets. 
To evaluate representations for classification, we select the following four datasets: the audio portion of Ryerson Audio-Visual Database of Emotional Speech and Song (\texttt{RAVDESS})~\cite{RAVDESS}, AudioMNIST (\texttt{AM})~\cite{AudioMNIST}, Spoken Language Understanding Resource Package (\texttt{SLURP})~\cite{SLURP}, and EMOVO~\cite{EMOVO} datasets.
Two datasets focus on emotion recognition (i.e., \texttt{RAVDESS} and \texttt{EMOVO}), whereas the remaining two address intent classification and digit recognition. 
All four speech datasets are single-label classification tasks, and their sample durations range from 0.6 to 3.7 seconds.
We utilize the provided training, validation, and test splits whenever available. Otherwise, we either implement cross-validation or create fixed partitions for all models tested in ARCH. 

\subsection{Model Architectures}
\label{sec:models}

The suite of SSL models evaluated in this work includes Wav2Vec 2.0 (\texttt{W2V2})~\cite{w2v2}, WavLM~\cite{wavlm}, HuBERT~\cite{hubert}, data2vec (\texttt{D2V})~\cite{data2vec}, and XLS-R~\cite{xlsr}.
All models leverage transformer architectures and are pre-trained on large amounts of unlabeled \textit{speech} data. 
Testing generalization beyond the original pre-training data distribution may limit accuracy in other domains, offering insights into how pre-training data impacts generalization capabilities.
All models operate directly on raw audio waveforms. 
A CNN front-end extracts frame-level features, which are then fed to the transformer encoder.
While other transformer-based SSL models operating on spectrograms have been proposed~\cite{koutini22passt, gong2022ssast}, we decided to focus on models operating on raw audio waveforms, leaving the evaluation of spectrogram-based approaches to future work.
The goal is to give insights into the selection of optimal domain-specific models, which may also serve as backbones for more complex architectures.

We evaluate the base (B), large (L), and extra-large (XL) versions of each model, with the number of parameters ranging from approximately 100M (B) to 300M (L) and 1B (XL).
Pre-training data includes LibriSpeech~\cite{librispeech}, Libri-Light~\cite{librilight}, GigaSpeech~\cite{chen21o_interspeech}, and VoxPopuli~\cite{voxpopuli}.
To address the lack of open-source pre-trained models for non-speech audio, we also release new models of different sizes that have been trained on general-purpose AudioSet~\cite{audioset} collection, enabling a more comprehensive evaluation of ARL models.

\section{Results and Analysis}
\label{sec:results}
Following a standard practice~\cite{HEAR}, we report the mean average precision (mAP) for multi-label classification tasks on FSD50K, MTT, and IRMAS datasets, and the accuracy for single-label classification tasks on all other datasets.
The results are summarized in Table \ref{tab:arch_complete}.

\begin{table*}[t!]
\centering
\caption{Performance of SSL models on ARCH benchmark. $^\diamond$ indicates models pre-trained on AudioSet. 
The best overall results are highlighted with a \colorbox[HTML]{DAE8FC}{light-blue} background and the best per model size are reported in \textbf{boldface}.}
\label{tab:arch_complete}
\vspace{0.5em}
\resizebox{0.98\linewidth}{!}{%
\begin{tabular}{@{}cccccc|cccc|cccc@{}}
\toprule
 &  & \multicolumn{4}{c|}{Acoustic Events} & \multicolumn{4}{c|}{Music} & \multicolumn{4}{c}{Speech} \\ \cmidrule(l){3-14} 
\multirow{-2}{*}{Model} & \multirow{-2}{*}{Size} & ESC-50 & US8K & FSD50K & VIVAE & FMA & MTT & IRMAS & MS-DB & RAVDESS & AM & SLURP & EMOVO \\ \midrule
W2V2 & \multicolumn{1}{c|}{B} & 45.73 & 55.48 & 19.39 & 31.47 & 50.54 & 37.56 & 35.14 & 66.06 & 55.32 & 86.38 & 14.37 & 31.80 \\
WavLM & \multicolumn{1}{c|}{B} & 49.88 & 61.84 & 17.63 & 36.31 & 48.71 & 34.93 & 32.62 & 54.18 & \textbf{67.94} & 99.50 & 30.98 & \textbf{43.08} \\
WavLM+ & \multicolumn{1}{c|}{B} & 58.73 & 64.07 & 21.57 & 36.17 & 56.17 & 38.24 & 35.76 & 57.51 & 52.20 & \textbf{99.63} & 28.06 & 36.73 \\
HuBERT & \multicolumn{1}{c|}{B} & 58.90 & 67.28 & 24.53 & \textbf{40.48} & 54.63 & 38.78 & 36.65 & 58.46 & 65.28 & 99.58 & 33.75 & 40.48 \\
D2V & \multicolumn{1}{c|}{B} & 23.63 & 45.63 & 10.06 & 30.19 & 40.58 & 27.60 & 25.87 & 50.74 & 48.03 & 99.06 & \textbf{43.57} & 27.27 \\
$^\diamond$W2V2-AS & \multicolumn{1}{c|}{B} & 52.61 & 70.48 & 21.29 & 31.26 & 59.50 & 37.92 & 35.85 & 64.61 & 45.94 & 88.09 & 11.00 & 30.83 \\
$^\diamond$HuBERT-AS & \multicolumn{1}{c|}{B} & \textbf{68.80} & \cellcolor[HTML]{DAE8FC}\textbf{79.09} & \textbf{31.05} & 40.06 & \textbf{65.87} & \textbf{43.44} & \textbf{47.67} & \textbf{67.81} & 63.54 & 98.84 & 20.53 & 33.39 \\ \midrule
W2V2 & \multicolumn{1}{c|}{L} & 13.13 & 42.70 & 5.80 & 22.01 & 41.71 & 20.95 & 19.91 & 50.23 & 11.57 & 45.74 & 7.33 & 19.27 \\
XLS-R & \multicolumn{1}{c|}{L} & 51.28 & 69.96 & 23.71 & 36.28 & 56.96 & 38.28 & 38.42 & 66.71 & 31.48 & 98.88 & 12.74 & 20.35 \\
WavLM & \multicolumn{1}{c|}{L} & 67.20 & 70.92 & 32.21 & 42.51 & 61.13 & 41.29 & 42.53 & 68.00 & 71.76 & 99.75 & 42.34 & \textbf{45.29} \\
HuBERT & \multicolumn{1}{c|}{L} & 63.98 & 70.00 & 29.51 & 40.95 & 54.79 & 38.36 & 36.81 & 64.08 & 72.57 & \cellcolor[HTML]{DAE8FC}\textbf{99.95} & \textbf{45.26} & 43.76 \\
D2V & \multicolumn{1}{c|}{L} & 25.35 & 49.15 & 10.82 & 30.57 & 43.46 & 28.52 & 27.08 & 44.20 & 45.14 & 99.15 & 28.60 & 23.07 \\
$^\diamond$W2V2-AS & \multicolumn{1}{c|}{L} & \cellcolor[HTML]{DAE8FC}\textbf{74.39} & \textbf{79.00} & \cellcolor[HTML]{DAE8FC}\textbf{37.58} & 39.65 & 66.58 & \cellcolor[HTML]{DAE8FC}\textbf{44.51} & 49.87 & 76.90 & 59.49 & 99.42 & 17.74 & 38.20 \\
$^\diamond$HuBERT-AS & \multicolumn{1}{c|}{L} & 71.52 & 75.63 & 37.41 & \cellcolor[HTML]{DAE8FC}\textbf{44.28} & \cellcolor[HTML]{DAE8FC}\textbf{67.54} & 43.35 & \cellcolor[HTML]{DAE8FC}\textbf{50.46} & \cellcolor[HTML]{DAE8FC}\textbf{77.82} & \textbf{73.26} & 99.59 & 20.46 & 38.61 \\ \midrule
XLS-R & \multicolumn{1}{c|}{XL} & \textbf{66.95} & \textbf{75.90} & \textbf{31.61} & 40.41 & \textbf{62.79} & \textbf{41.99} & \textbf{43.57} & \textbf{69.79} & 55.44 & 99.86 & 25.14 & 34.58 \\
HuBERT & \multicolumn{1}{c|}{XL} & 63.40 & 69.66 & 29.32 & \textbf{42.72} & 56.25 & 37.76 & 37.30 & 64.71 & \cellcolor[HTML]{DAE8FC}\textbf{75.69} & \cellcolor[HTML]{DAE8FC}\textbf{99.95} & \cellcolor[HTML]{DAE8FC}\textbf{47.81} & \cellcolor[HTML]{DAE8FC}\textbf{47.17} \\ \bottomrule
\end{tabular}%
}
\end{table*}

\subsection{Acoustic Events}

On the acoustic events datasets, the speech-pretrained models demonstrate reasonable generalization capabilities. 
Among the base models, HuBERT and WavLM+ achieve the top performance on most datasets, with HuBERT showing a consistent advantage. 
However, the impact of pre-training data distribution becomes evident while comparing speech-pretrained HuBERT to our HuBERT-AS model trained on AudioSet. 
HuBERT-AS substantially outperforms HuBERT on 3 out of 4 datasets, highlighting the benefits of wide-ranging pre-training data encompassing multiple audio domains.
Further, increasing model size from base to large provides additional gains, Wav2Vec 2.0 pre-trained on AudioSet (W2V2-AS) achieves the highest accuracy on 3 out of 4 datasets, closely followed by HuBERT-AS.
This demonstrates the combined advantages of expanded model capacity and in-domain pre-training for acoustic events tasks.

\noindent
For extra-large models, we only evaluate the speech-pretrained versions given the computational demands. 
XLS-R shows some improvement over its large counterpart (possibly due to the more varied pre-training data), but scaling up HuBERT does not yield any significant improvement.
While speech pre-training can generalize well to acoustic events, AudioSet pre-training substantially improves the system performance. 
Also, a larger model scale enables the new AudioSet-pretrained models to achieve top performance. 

\subsection{Music}

In the music domain, our AudioSet-pretrained models achieve the best performance. 
The base HuBERT-AS model outperforms all other base models, and the large HuBERT-AS model achieves the highest accuracy on 3 out of 4 datasets, with the exception of \texttt{MTT} where W2V2-AS performs slightly better.
Similar to the audio events domain, increasing model size consistently improves performance, but even extra-large speech-pretrained models fail to surpass the AudioSet models, highlighting the importance of diverse pre-training data for music tasks.
The gains achieved by scaling model size are evident on the instrument detection task of \texttt{MS-DB}, with over 14\% improvement from base to large models pre-trained on AudioSet.
Interestingly, HuBERT-AS on average outperforms W2V2-AS in both base and large sizes, suggesting that incorporating discrete targets during pre-training provides advantages for learning musically-relevant representations.
Similar to the acoustic events domain, diversifying pre-training data provides significant benefits for music tasks.

\subsection{Speech}

As expected, the speech-pretrained models achieve the highest performance on the speech domain, clearly outperforming the AudioSet models given the in-domain pre-training data.
Scaling up model size provides consistent and often substantial gains (e.g., +8\% on \texttt{RAVDESS} from best-performing base to large models).
The extra-large HuBERT model achieves the highest accuracy across all speech datasets, highlighting the importance of model capacity for this domain.
Among the base models, WavLM performs averagely best, leveraging advantages from its masked prediction pre-training task. 
For large models, both WavLM and HuBERT clearly show superior performance, indicating the benefits of discrete targets at a greater scale.
Interestingly, considering the performance of extra-large models on the Italian emotion recognition dataset \texttt{EMOVO}, English-only pre-trained HuBERT achieves the top score, significantly exceeding even XLS-R, which leverages cross-lingual pre-training data. 
This suggests HuBERT learns more generalizable representations of speech.
Larger models enable significant improvements, with 
HuBERT-XL showing dominant performance due to its scale and training approach.

\section{Discussion}
Through extensive analysis on ARCH using different SSL models, several key insights emerge.
First, pre-training models with heterogeneous training data provide significant benefits for non-speech tasks, highlighting the importance of pre-training data to learn broadly applicable representations. 
Our work confirms the generalizability of Transformer models operating on raw audio and their SSL pre-training objectives also for non-speech audio domains.

HuBERT-based models achieve the highest performance overall, demonstrating the advantages of pre-training with discrete targets.
In nearly all cases, increasing model size from base to large consistently improves performance.
However, the analysis also reveals that model capacity is not saturated even for base-sized models.
More data leads to more transferable learning, even when the additional data comes from a single domain.
For example, comparing \texttt{WavLM} and \texttt{WavLM+} shows that despite identical architectures and pre-training objectives, the additional speech data used to train \texttt{WavLM+} results in more generally capable representations.
Optimizing pre-training objectives and data diversity will be key to developing more general and effective representations.

\vspace{1mm}
\noindent
\textbf{Limitations} While the evaluation of several ARL models has provided valuable insights, the findings reported in this work are limited by the scope of the benchmark.
This work focuses on representations extracted from raw audio waveforms, but future efforts will also evaluate spectrogram-based approaches. 
The evaluation of extra-large models pre-trained on general audio datasets was limited by computational resources, restricting the analysis to speech-pretrained models. 
While the potential of speech pre-training has been already demonstrated in other downstream tasks~\cite{SUPERB}, further evaluations of newly released models on other tasks (e.g., audio captioning) are advisable to fully understand the benefits of diverse pre-training data.

\section{Conclusion} 
In this work, we introduced ARCH, a new benchmark for audio representation learning, and performed an extensive comparative analysis.
The results demonstrate the benefits of pre-training on large, multi-domain datasets for learning widely useful representations. 
This highlights the need to develop more extensive general audio datasets~\cite{lee2021acav100m}.
While increasing model size has consistently improved results, further optimizing pre-training data and objectives remains critical for learning cross-domain representations.
We believe releasing new open-source models and standardizing evaluation through ARCH will accelerate progress in evaluating audio representation learning models.

\clearpage

\bibliographystyle{IEEEbib}
{\footnotesize
\bibliography{refs}
}

\end{document}